\documentstyle[12pt,a4,epsf]{article}

\def\tr{\mathop{\rm tr}\nolimits}
\def\Tr{\mathop{\rm Tr}\nolimits}

\makeatletter
\@addtoreset{equation}{section}
\makeatother

\newcommand{\VEV}[1]{\left\langle #1 \right\rangle}

\newcommand{\nn}{\nonumber}

\newcommand{\bequ}{\begin{equation}}
\newcommand{\eequ}{\end{equation}}
\newcommand{\beqn}{\begin{eqnarray}}
\newcommand{\eeqn}{\end{eqnarray}}

\begin{document}
\begin{titlepage}

\begin{flushright}
hep-ph/0204030\\
KUNS-1777\\
BA-02-15\\
\today
\end{flushright}

\vspace{4ex}

\begin{center}
{\large \bf
Supersymmetric $SU(3)^3$ unification with Anomalous $U(1)_A$ Gauge Symmetry
}

\vspace{6ex}

\renewcommand{\thefootnote}{\alph{footnote}}
Nobuhiro Maekawa\footnote
{e-mail: maekawa@gauge.scphys.kyoto-u.ac.jp
}
and 
Qaisar Shafi\footnote{
email: shafi@bartol.udel.edu
}

\vspace{4ex}
${}^{a}${\it Department of Physics, Kyoto University, Kyoto 606-8502, Japan}\\
${}^{b}${\it Bartol Research Institute, University of Delaware, Newark,
DE19716, USA}\\
\end{center}

\renewcommand{\thefootnote}{\arabic{footnote}}
\setcounter{footnote}{0}
\vspace{6ex}

\begin{abstract}
We consider supersymmetric unification based on the gauge symmetry
$SU(3)_C\times SU(3)_L\times SU(3)_R$ supplemented by an anomalous
$U(1)_A$ gauge symmetry. Realistic fermion masses and mixings are 
realized, including large mixings in the neutrino sector. We also
consider the supersymmetric flavor problem, gauge coupling
unification and proton decay. The dominant proton decay mode is
expected to be $p\rightarrow e^+\pi^0$ and the lifetime is estimated
to be $\sim 10^{34}-10^{35}$ years.

\end{abstract}

\end{titlepage}

\section{Introduction}
 Supersymmetric grand unified theories offer a particularly elegant scenario
for unifying the strong and electroweak interactions. Recently, realistic
models based on $SO(10)$\cite{maekawa,maekawa2,maekawa3} and 
$E_6$\cite{BM,MY} have been proposed in which an anomalous $U(1)_A$ 
gauge symmetry\cite{U(1)} plays a critical role. The anomaly is
cancelled via the Green Schwarz mecanism\cite{GS}, and the resulting 
phenomenology has several attractive features. In particular, all 
interactions allowed by the symmetries are included in the discussion 
(there can be undetermined order unity coefficients accompanying the 
interactions. These models naturally resolve the doublet- triplet 
problem \cite{DTsplitting,extra} using the mechanism of
reference\cite{DW} (see also\cite{BarrRaby,Chako,complicate}). 
Realistic pattern of qaurk and lepton mass
matrices, including large neutrino mixings\cite{SK}, are realized, 
using the Froggatt- Nielsen (FN) mechanism\cite{FN}. The anomalous 
$U(1)_A$ also helps explain the hierarchical symmetry breaking scales 
and the masses acquired by the superheavy particles, and leads 
precisely to the minimal supersymmetric standard model (MSSM) at 
low energies. Even though the gauge couplings unify in these schemes 
slightly below the usual GUT scale $2\times10^{16}$ GeV, dimension
five proton decay is sufficiently suppressed , and the decay 
$p--> e +\pi$ via gauge mediated dimension six operators may be seen 
in the near future.  
Finally, in these models the cutoff scale is lower than the Planck scale
$M_{Planck}$ and the $\mu$ problem is also resolved.

However, these models require two adjoint Higgs fields to realize
DT splitting, which is not so easily realized 
in the framework of superstring models. In this paper we examine the
application of the above approach to grand unified theories with
semi-simple unification whose symmetry breaking to MSSM (the minimal
supersymmetric standard model) does not require adjoint Higgs field.
A particularly attractive example is provided by the gauge symmetry
$SU(3)^3\equiv SU(3)_C\times SU(3)_L\times SU(3)_R$ which is a maximal
subgroup of $E_6$, and which arises as an effective four dimensional
symmetry from the compactification of the $E_8\times E_8$ heterotic
superstring theory on a Calabi-Yau manifold\cite{GSW}.
Phenomenology based on $SU(3)^3$ has been extensively discussed in the
past\cite{shafi1,shafi2,shafi3}. Our goal here is to apply the techniques
of \cite{maekawa}-\cite{MY} to gauge symmetry $SU(3)^3$ and elucidate
the most important consequences. In particular, we will show how realistic
fermion masses and mixings are obtained, including large mixings in the
neutrino sector (with the exception of $U_{e3}$). 
We also consider SUSY breaking and flavor changing neutral
currents, gauge coupling unification and proton decay. While dimension
five proton decay turns out to be heavily suppressed, dimension six operators
yield a proton lifetime of $\sim 10^{34}-10^{35}$ years for 
the decay channel $p\rightarrow e^+\pi^0$.

\section{Matter sector}
The matter sector has essentially the same structure as the $E_6$ model
\cite{BM}, with the ${\bf 27}$ of $E_6$ given in terms of
$SU(3)_C\times SU(3)_L\times SU(3)_R$ as
\begin{equation}
{\bf 27}\rightarrow ({\bf 3},{\bf \bar 3},{\bf 1})+
({\bf \bar 3},{\bf 1},{\bf 3})+({\bf 1},{\bf 3},{\bf \bar 3}).
\end{equation}
Three ${\bf 27}$-plets $\Psi_i$ ($i=1,2,3$) are introduced, 
and the Yukawa interations contain appropriate powers 
of the VEV of FN field $\VEV{\Theta}=\lambda \Lambda$, 
which has an anomalous $U(1)_A$ charge $\theta=-1$, namely
\begin{equation}
\lambda^{\psi_i+\psi_j+\phi}\Psi_i\Psi_j\Phi.
\end{equation}
Here $\Phi$ is a Higgs field, and 
throughout this paper we use units in which the cutoff $\Lambda=1$, and
denote all
superfields by uppercase letters and their anomalous $U(1)_A$ charges
by the corresponding lowercase letters.
Using the definitions of the fields $Q({\bf 3,2})_{\frac{1}{6}}$,
$U^c({\bf \bar 3,1})_{-\frac{2}{3}}$, $D^c({\bf \bar 3,1})_{\frac{1}{3}}$,
$L({\bf 1,2})_{-\frac{1}{2}}$, $E^c({\bf 1,1})_1,N^c({\bf 1,1})_0$,
$L^{c\prime}({\bf 1,2})_{\frac{1}{2}}$, $L'({\bf 1,2})_{-\frac{1}{2}}$, 
$D'({\bf 3,1})_{-\frac{1}{3}}$, $D^{c\prime}({\bf \bar 3,1})_{\frac{1}{3}}$,
$S({\bf 1,1})$, and their conjugate fields under the standard
model (SM) gauge symmetry, the fields 
$({\bf 3},{\bf \bar 3},{\bf 1})$, 
$({\bf \bar 3},{\bf 1},{\bf 3})$, and 
$({\bf 1},{\bf  3},{\bf \bar 3})$ under 
$SU(3)_C\times SU(3)_L\times SU(3)_R$ are 
\begin{eqnarray}
({\bf 3},{\bf \bar 3},{\bf 1})&\rightarrow & Q+D' \\
({\bf \bar 3},{\bf 1},{\bf 3})&\rightarrow & U^c+D^c+D^{c\prime} \\
({\bf 1},{\bf 3},{\bf \bar 3})&\rightarrow & L+E^c+N^c+L^{c\prime}+L'+S.
\end{eqnarray}
For future reference, under the breaking $E_6\rightarrow SO(10)$,
\begin{equation}
{\bf 27}\rightarrow 
\underbrace{[Q+U^c+E^c+D^c+L+N^c]}_{\bf 16}
+\underbrace{[L^{c\prime}+L'+D'+D^{c\prime}]}_{\bf 10}
+\underbrace{S}_{\bf 1}.
\end{equation}
Since $D'(L^{c\prime})$ can acquire superheavy mass by combining
with a linear combination of $D^c$ and $D^{c\prime}$ ($L$ and $L'$)
after the breaking $SU(3)_C\times SU(3)_L\times SU(3)_R$
to the SM gauge group, 
the remaining massless fields form the three generation matter content
of MSSM.
Since the Yukawa couplings are determined mainly by the anomalous 
$U(1)_A$ charges
of the massless fields, we would like to know which of the fields among 
$D^c_i$ and $D^{c\prime}_i$ ($L_i$ and $L'_i$) ($i=1,2,3$)
are massless. 

To this end, we discuss how  
$SU(3)_C\times SU(3)_L\times SU(3)_R$ breaks to MSSM.
We introduce the Higgs fields with non-vanishing VEVs
$\Phi({\bf 1},{\bf 3},{\bf \bar 3})$, 
$\bar \Phi({\bf 1},{\bf \bar 3},{\bf 3})$,
$C({\bf 1},{\bf 3},{\bf \bar 3})$, and
$\bar C({\bf 1},{\bf \bar 3},{\bf 3})$.
The VEVs 
$|\VEV{\Phi}|=|\VEV{\bar \Phi}|\sim \lambda^{-(\phi+\bar \phi)/2}$ 
break
$SU(3)_C\times SU(3)_L\times SU(3)_R$ to
$SU(3)_C\times SU(2)_L\times SU(2)_R\times U(1)_{B-L}$,
while 
$|\VEV{C}|=|\VEV{\bar C}|\sim \lambda^{-(c+\bar c)/2}$ 
break
$SU(3)_C\times SU(2)_L\times SU(2)_R\times U(1)_{B-L}$
to the SM gauge group.
Here the VEVs are determined by the anomalous $U(1)_A$ charges,
and the reason is roughly as follows
(We will explain how to determine the VEVs later):
\begin{enumerate}
\item
Since the interactions are determined by the anomalous $U(1)_A$ charges,
the VEV of the gauge invariant operator $O$ with negative charge $o$
is determined as $\VEV{O}\sim \lambda^{-o}$.
\item
For $O=\bar\Phi\Phi$, the VEV becomes 
$\VEV{\bar\Phi\Phi}\sim \lambda^{-(\phi+\bar\phi)}$.
\item
Since the $D$-flatness condition requires 
$|\VEV{\bar \Phi}|=|\VEV{\Phi}|$,
we obtain
$|\VEV{\bar \Phi}|=|\VEV{\Phi}|\sim \lambda^{-\frac{1}{2}(\bar\phi+\phi)}$.
\end{enumerate}
The massless modes can be determined from the superpotential
\begin{equation}
W_1=\lambda^{\psi_i+\psi_j+\phi}\Psi_i\Psi_j\Phi
+\lambda^{\psi_i+\psi_j+c}\Psi_i\Psi_jC,
\end{equation}
and the VEVs 
$\VEV{\Phi}\sim \lambda^{-\frac{1}{2}(\phi+\bar \phi)}$ and
$\VEV{C}\sim \lambda^{-\frac{1}{2}(c+\bar c)}$.
Here, for simplicity, we have assumed the $E_6$ like charge assignment
in the matter sector, but in principle, we can assign these charges
without respecting $E_6$ symmetry.
The mass matrices of $D^{c(\prime)}$ and $D'$ ($L^{(\prime)}$ and 
$L^{c\prime}$) are
obtained from
\begin{equation}
M_I=\bordermatrix{
&I_1&I_2&I_3&I'_1&I'_2&I'_3\cr
I^{c\prime}_1&\lambda^{2\psi_1+r}&\lambda^{\psi_1+\psi_2+r}
&\lambda^{\psi_1+\psi_3+r}
&\lambda^{2\psi_1}&\lambda^{\psi_1+\psi_2}
&\lambda^{\psi_1+\psi_3}  \cr
\bar I'_2&\lambda^{\psi_1+\psi_2+r}&\lambda^{2\psi_2+r}
&\lambda^{\psi_2+\psi_3+r}
& \lambda^{\psi_1+\psi_2}   &  \lambda^{2\psi_2}
&\lambda^{\psi_2+\psi_3} \cr
\bar I'_3 &\lambda^{\psi_1+\psi_3+r}&\lambda^{\psi_2+\psi_3+r}
&\lambda^{2\psi_3+r}
 &\lambda^{\psi_1+\psi_3} & \lambda^{\psi_2+\psi_3}
 & \lambda^{2\psi_3}  \cr} \lambda^{\frac{1}{2}(\phi-\bar\phi)},
 \label{full}
\end{equation}
where $I=L, D^c$ and we have defined  the parameter $r$ as 
\begin{equation}
r\equiv \frac{1}{2}(c-\bar c-\phi+\bar\phi),\label{r}
\end{equation}
which we use frequently in the following discussion. 
Note that the mass matrices are determined by the anomalous $U(1)_A$ charges.
Therefore the massless modes are also determined by the charges.
As discussed in Ref.\cite{BM}, as long as we neglect the cases with 
vanishing coefficients arising from some SUSY based mechanism,
the main components of the massless modes can be obtained as follows:
\begin{enumerate}
\item  $\psi_1-\psi_3<r: (I_1,I_2,I_3)$.
\item  $0<r<\psi_1-\psi_3: (I_1,I'_1,I_2)$.
\item  $\psi_3-\psi_1<r<0: (I_1,I'_1,I'_2)$.
\item  $r<\psi_3-\psi_1: (I'_1,I'_2,I'_3)$.
\end{enumerate}
The case $(I_1,I'_1,I_2)$ is interesting, because bi-large neutrino mixing
angles can be realized without $\tan \beta$ too small, if we take account of
the mixing of subcomponents. Indeed, the massless modes $(I_1^0,I_2^0,I_3^0)$
are given by
\begin{eqnarray}
I_1^0 &=& I_1
+\lambda^{\psi_1-\psi_3}I_3
+\lambda^{\psi_1-\psi_2+r}I'_2
+\lambda^{\psi_1-\psi_3+r}I'_3, 
\label{51} \\
I^0_2 &=& I'_1+\lambda^{\psi_1-\psi_3-r}I_3
+\lambda^{\psi_1-\psi_2}I'_2+\lambda^{\psi_1-\psi_3}I'_3, 
\label{52} \\
I^0_3 &=& I_2+\lambda^{\psi_2-\psi_3}I_3
+\lambda^{r}I'_2+\lambda^{\psi_2-\psi_3+r}I'_3,
\label{53}
\end{eqnarray}
where the first terms on the right-hand sides are the main components of
these massless modes, and the other terms represent mixing with
the other states, $I_3$, $I'_2$ and $I'_3$.

The mass matrices for quarks and leptons are obtained from the 
superpotential
\begin{equation}
W_2=\lambda^{\psi_i+\psi_j+h}\Psi_i\Psi_jH,\label{yukawa}
\end{equation}
where $H({\bf 1},{\bf 3},{\bf \bar 3})$ contains the MSSM Higgs doublets.
If we adopt the charges $\psi_1=3+n$, $\psi_2=2+n$, $\psi_1=n$, and $h=-2n$,
the mass matrices
are given by
\begin{eqnarray}
M_U &=&
\bordermatrix{
   &   U^c_1  &  U^c_2    &     U^c_3 \cr
Q_1&\lambda^6 & \lambda^5 & \lambda^3 \cr
Q_2&\lambda^5 & \lambda^4 & \lambda^2 \cr
Q_3&\lambda^3 & \lambda^2 & 1 \cr}
\VEV{L^{c\prime}(H)},\\
M_D(M^T_E\eta^{-1}) &=&\bordermatrix{
           &  D^{c0}_1(L_1^0)  &   D^{c0}_2(L_2^0)  &   D^{c0}_3(L^0_3) \cr
Q_1(E^c_1) &  \lambda^6        &  \lambda^{6-r}     & \lambda^5  \cr
Q_2(E^c_2) &  \lambda^5        &  \lambda^{5-r}     & \lambda^4 \cr
Q_3(E^c_3) &  \lambda^3        &  \lambda^{3-r}     & \lambda^2
 \cr}\VEV{L'(H)}.
\end{eqnarray}
Here $\eta\sim 2-3$ is the renormalization group factor.
Then we can obtain the CKM matrix
as
\begin{equation}
U_{\rm CKM}=
\left(
\begin{array}{ccc}
1 & \lambda &  \lambda^3 \\
\lambda & 1 & \lambda^2 \\
\lambda^3 & \lambda^2 & 1
\end{array}
\right),
\label{CKM}
\end{equation}
which reproduces the experimental results if we take
$\lambda\sim 0.2$.
Since the ratio of the Yukawa couplings of the top and bottom quarks is
$\lambda^2$,
a small value of 
$\tan \beta
\sim m_t/m_b\cdot \lambda^2$ is
predicted
by these mass matrices.

The Dirac neutrino mass matrix is given by the $3\times 6$ matrix
\begin{equation}
\bordermatrix{
    &S_1&S_2&S_3&N^c_1&N^c_2 &N^c_3   \cr
L^0_1 &  \lambda^{r+6}  &  \lambda^{r+5}   & \lambda^{r+3}  
&  \lambda^6 &  \lambda^5     & \lambda^3  \cr
L^0_2  &  \lambda^6   &  \lambda^5  &\lambda^3 &  
\lambda^{6-r} &  \lambda^{5-r}&\lambda^{3-r}\cr
L^0_3  & \lambda^{r+5}   &  \lambda^{r+4}  & \lambda^{r+2}
&  \lambda^5   &\lambda^4  &\lambda^2
 \cr}\VEV{L^{c\prime}(H)}\eta,
\end{equation}
which we simply express as
\begin{equation}
M_{N}=
\left(
\begin{array}{cc}
 \lambda^{r+2} & \lambda^2
\end{array}
\right)\otimes
\left(
\begin{array}{ccc}
\lambda^4 & \lambda^3 & \lambda \\
\lambda^{4-r} & \lambda^{3-r} & \lambda^{1-r} \\
\lambda^3   & \lambda^2         & 1
\end{array}
\right)\VEV{L^{c\prime}(H)}\eta.
\end{equation}
The $6\times 6$ matrix for the right-handed neutrinos
($S_i, i=1,2,3$, and $N^c_k ,k=1,2,3$ )
is obtained as 
\begin{eqnarray}
M_R&=
&\lambda^{\psi_i+\psi_j+2\bar \phi}S_iS_j\VEV{\bar \Phi}^2
+\lambda^{\psi_i+\psi_k+\bar c+\bar \phi}S_iN^c_k
\VEV{\bar \Phi}\VEV{\bar C}, \nonumber \\
&&+\lambda^{\psi_k+\psi_m+2\bar c}N^c_kN^c_m\VEV{\bar C}^2 \\
&=&\lambda^{2n}
\left(
\begin{array}{cc}
\lambda^{\bar \phi-\phi} & \lambda^{(\bar \phi-\phi+\bar c-c)/2} \\
\lambda^{(\bar \phi-\phi+\bar c-c)/2} & \lambda^{\bar c-c}
\end{array}
\right)
\otimes
\left(
\begin{array}{ccc}
\lambda^6 & \lambda^5 & \lambda^3 \\
\lambda^5 & \lambda^4 & \lambda^2 \\
\lambda^3   & \lambda^2         & 1
\end{array}
\right)
\end{eqnarray}
from the interactions
\begin{equation}
\lambda^{\psi_i+\psi_j+2\bar \phi}\Psi_i\Psi_j\bar \Phi\bar \Phi
+\lambda^{\psi_i+\psi_j+\bar c+\bar \phi}\Psi_i\Psi_j\bar \Phi\bar C
+\lambda^{\psi_i+\psi_j+2\bar c}\Psi_i\Psi_j\bar C\bar C,
\end{equation}
by developing the VEVs 
$\VEV{\bar \Phi}\sim \lambda^{-\frac{1}{2}(\phi+\bar \phi)}$ and
$\VEV{\bar C}\sim \lambda^{-\frac{1}{2}(c+\bar c)}$.
Using the seesaw mechanism
\cite{seesaw},
we obtain the neutrino mass matrix
\begin{equation}
M_\nu=M_NM_R^{-1}M_N^T=\lambda^{4-2n+c-\bar c}\left(
\begin{array}{ccc}
\lambda^2 & \lambda^{2-r} & \lambda \\
\lambda^{2-r} & \lambda^{2-2r} & \lambda^{1-r} \\
\lambda   & \lambda^{1-r}         & 1
\end{array}
\right)\VEV{L^{c\prime}(H)}^2\eta^2.
\end{equation}
Then, we finally
obtain the Maki-Nakagawa-Sakata matrix as
\begin{equation}
U_{\rm MNS}=
\left(
\begin{array}{ccc}
1 & \lambda^r &  \lambda \\
\lambda^{r} & 1 & \lambda^{1-r} \\
\lambda & \lambda^{1-r} & 1
\end{array}
\right).
\end{equation}

If we adopt $r=1/2$, namely,
\begin{equation}
c-\bar c=\phi-\bar \phi+1,
\label{rr}
\end{equation}
bi-large neutrino mixing angles are realized because of
$\lambda^{1/2}\sim 0.5$ as within the
$SO(10)$ model in Ref. \cite{maekawa} and $E_6$ model in
Ref. \cite{BM}.
Moreover, it predicts
$U_{e3}\sim \lambda$. 
Future experiments can see whether there is 
such a large $U_{e3}$ just below the  CHOOZ  upper
limit $U_{e3}\leq 0.15$
\cite{CHOOZ} or not.
For the neutrino masses, the model predicts
$m_{\nu_\mu}/m_{\nu_\tau}\sim \lambda$,
which is consistent with
the most probable
LMA MSW solution for the solar neutrino puzzle.
\cite{SK}

If we define
\begin{equation}
l\equiv \bar \phi-\phi+2n-10,
\end{equation}
the neutrino mass matrix is given by
\begin{equation}
M_\nu=\lambda^{-(5+l)}\left(
\begin{array}{ccc}
\lambda^2 & \lambda^{1.5} & \lambda \\
\lambda^{1.5} & \lambda & \lambda^{0.5} \\
\lambda   & \lambda^{0.5}         & 1
\end{array}
\right)\VEV{L^{c\prime}(H)}^2\eta^2,
\end{equation}
where we have used the relation (\ref{rr}). 
The paramter $l$ can be determined by
\begin{equation}
\lambda^l=\lambda^{-5}\frac{\VEV{L^{c\prime}(H)}^2\eta^2}
{m_{\nu_\tau}\Lambda}.
\end{equation}
We are supposing that the cutoff scale $\Lambda$ is in the range
$5\times 10^{15}{\rm GeV}<\Lambda<10^{20}{\rm GeV}$, which allows
$-3\leq l \leq 2$. 
If we choose $l=-2$,
the neutrino masses are given  by
$m_{\nu_\tau}\sim \lambda^{-3}\VEV{L^{c\prime}(H)}^2\eta^2/\Lambda\sim
m_{\nu_\mu}/\lambda
\sim m_{\nu_e}/\lambda^2$. If we take $\eta\VEV{L^{c\prime}(H)}=100$ GeV,
$\Lambda\sim 2\times 10^{16}$ GeV and $\lambda=0.2$, then we get
$m_{\nu_\tau}\sim 6\times 10^{-2}$ eV, $m_{\nu_\mu}\sim 1\times
10^{-2}$ eV
and $m_{\nu_e}\sim 2\times 10^{-3}$ eV. 
These values are pretty much consistent with 
the experimental data for atmospheric neutrinos and 
with the large mixing angle (LMA) MSW solution for the solar neutrino 
problem.\cite{MSW}

\section{SUSY breaking and FCNC}
Let us now discuss SUSY breaking. Since the
anomalous $U(1)_A$ charges depend on flavor to produce the
hierarchy of Yukawa couplings, generically
non-degenerate scalar fermion masses are induced through the anomalous
$U(1)_A$ $D$-term. 
Under the $E_6$-like charge assignment of the matter sector, the
SUSY contribution to $K^0-\bar K^0$ mixing is naturally suppressed as
in $E_6$ GUT scenario.
The essential point is that the anomalous $U(1)_A$ charge of $D_1^{c0}$
becomes the same as that of $D_2^{c0}$, because the field 
$D_1^{c0}\sim D_1^c$ and $D_2^{c0}\sim D_1^{c\prime}$ arise from
a single field $\Psi_1$ of $E_6$. Since the constraints from $K^0-\bar K^0$ 
mixing to the ratio 
$\delta\equiv \Delta/\tilde m^2$, where
$\Delta$ is the mixing of sfermion mass matrices and 
$\tilde m$ is the average of sfermion mass,
are
\begin{eqnarray}
&&\sqrt{|{\rm Re} (\delta_{12}^D)_{LL}(\delta_{12}^D)_{RR}|}
\leq 2.8\times 10^{-3}
\left( \frac{\tilde m_q ({\rm GeV})}{500}\right), \label{LR}\\
 &&|{\rm Re} (\delta_{12}^D)_{LL}|,|{\rm Re} (\delta_{12}^D)_{RR}|
 \leq 4.0\times 10^{-2}
\left( \frac{\tilde m_q ({\rm GeV})}{500}\right), \label{LL}
\end{eqnarray}
and the former constraint is much stronger than the latter constraint,
suppression of $(\delta_{12}^D)_{RR}$ makes the constraint
on the SUSY breaking sector much weaker. 
Here we use the notation 
$(\delta_{ij}^F)_{XY}$, where
$F=U,D,N,E$, the chirality index is $X,Y=L,R$, 
and the generation index is $i,j=1,2,3$,
as defined in Ref. \cite{masiero}.
As in the usual anomalous $U(1)_A$ scenario, $\Delta$ can be
estimated as
\begin{equation}
(\Delta_{ij}^F)_{XX}\sim
\lambda^{|f_i-f_j|}(|f_i-f_j|)\VEV{D_A}.
\end{equation}
Therefore, the coincidence of the anomalous $U(1)_A$ charges 
of $D^{c0}_1$ and $D^{c0}_2$ leads to the suppression of 
$(\Delta_{12}^D)_{RR}$.
This  weakens the constraints from $K^0-\bar K^0$ mixing.

To see how weak the constraints become in our scenario, we
fix the SUSY breaking sector as follows.
At the cutoff scale, we adopt the common gaugino mass $M_{1/2}$ and
$D$-term of anomalous $U(1)_A$ gauge symmetry $D_A$ as 
non-vanishing SUSY breaking parameters.
Then, the scalar fermion mass squared at low energy scales is estimated as
\begin{equation}
\tilde m_{F_i}^2\sim f_i R M_{1/2}^2+\eta_FM_{1/2}^2,
\end{equation}
where $\eta_F$ is a renormalization group factor and
\begin{equation}
R\equiv \frac{\VEV{D_A}}{M_{1/2}^2}.
\end{equation}
The constraint (\ref{LL}) for 
$(\delta_{12}^D)_{LL}$ is rewritten 
\begin{equation}
M_{1/2}\geq 
1.25\times 10^4 \lambda\frac{(\psi_1-\psi_2)R}
{(\eta_{D_L}+\frac{\psi_1+\psi_2}{2}R)^{3/2}}
({\rm GeV}).
\end{equation}
Though the main contribution to $(\delta_{12}^D)_{RR}$ vanishes, through
the mixing in Eqs. (\ref{51}) and (\ref{52}), $(\delta_{12}^D)_{RR}$ is
estimated as
\begin{equation}
(\delta_{12}^D)_{RR}\sim \lambda^{\frac{1}{2}}\frac{\lambda^2(\psi_1-\psi_2)R}
                                                  {\eta_{D_R}+\psi_1R}.
\label{RR}
\end{equation}
From Eq. (\ref{LR}) for $\sqrt{(\delta_{12}^D)_{LL}(\delta_{12}^D)_{RR}}$,
the constraint on the gaugino mass $M_{1/2}$ is given by
\begin{equation}
M_{1/2}\geq 1.8\times 10^5\frac{\lambda^{1.75} R(\psi_1-\psi_2)}
                               {(\eta_D +\psi_1 R)^{1.5}}.
\end{equation}

On the other hand,
the $\mu\rightarrow e \gamma$ process gives
\begin{equation}
\ |(\delta_{12}^E)_{LL}|,|(\delta_{12}^E)_{RR}|\leq 3.8\times 10^{-3}
\left( \frac{\tilde m_l ({\rm GeV})}{100}\right)^2,\label{LLl}
\label{LFV}
\end{equation}
where $\tilde m_l$ is the average mass of the scalar leptons.
This constraint is rewritten
\begin{equation}
M_{1/2}\geq 1.6\times 10^3
\frac{(\lambda (\psi_1-\psi_2)R)^{1/2}}
{\eta_{E_R}+\frac{\psi_1+\psi_2}{2}R}
({\rm GeV}).
\end{equation}

Taking the values $\psi_1=9/2$, $\psi_2=7/2$, 
$\eta_{D_L}\sim \eta_{D_R}\sim 6$ and 
$\eta_{E_R}\sim 0.15$, 
the rough lower limits on the gaugino mass are in
Table I.

\vspace{3mm}
\begin{center}
Table I. Lower bound on gaugino mass $M_{1/2}$ at the GUT scale (in GeV).
\begin{tabular}{|c|c|c|c|c|c|} 
\hline
$R$                              & 0.1  & 0.3 & 0.5 & 1 & 2 \\ \hline
$(\delta_{12}^D)_{LL}$   & 17 & 43  & 61  & 87 & 105 \\ \hline
$\sqrt{(\delta_{12}^D)_{LL}(\delta_{12}^D)_{RR}}$   
                         & 78 & 191 & 268 & 373 & 437 \\ \hline
$|(\delta_{12}^E)_{RR}| $          & 431 & 304 & 221 & 161 & 116\\ \hline
\end{tabular}

\vspace{5mm}
\end{center}
Note that in some range of $R$, the $\mu\rightarrow e\gamma$
process gives the severest constraint among the FCNC 
processes.\cite{kurosawa}
Therefore, the lepton flavour violating processes\cite{kurosawa,LFV} 
might be seen in the near future.

\section{Higgs sector}
In addition to the Higgs with non-vanishing VEVs $\Phi$, $\bar \Phi$,
$C$, and $\bar C$, we introduce $C'({\bf 1},{\bf 3},{\bf \bar 3})$, 
$\bar C'({\bf 1}, {\bf \bar 3},{\bf 3})$ with vanishing VEVs 
and several singlets $S$ and $Z$ 
in order to give superheavy masses to these Higgs fields. 
The Higgs content is
\vspace{3mm}
\begin{center}
Table II. The typical values of anomalous $U(1)_A$ charges are listed.

\begin{tabular}{|c|c|c|} 
\hline
                  &   non-vanishing VEV  & vanishing VEV \\
\hline 
$({\bf 1},{\bf 3},{\bf \bar 3})$
                  &   $\Phi(\phi=-3)$\  $C(c=-3)$ &  $C'(c'=0)$  \\
$({\bf 1},{\bf \bar 3},{\bf 3})$
                  & $\bar \Phi(\bar \phi=2)$ \  $\bar C(\bar c=1)$ &
                  $\bar C'(\bar c'=4)$ \\
{\bf 1}           &   $Z_i(z_i=-1)(i=1,2,3)$  &  \\
$({\bf 3},{\bf \bar 3},{\bf 1})$ &   & $Q_L(q_l=1)$  \\
$({\bf \bar 3},{\bf 3},{\bf 1})$ &   & $\bar Q_L(\bar q_l=0)$  \\
$({\bf \bar 3},{\bf 1},{\bf 3})$ &   & $Q_R(q_r=1)$  \\
$({\bf 3},{\bf 1},{\bf \bar 3})$ &   & $\bar Q_R(\bar q_r=0)$  \\
\hline
\end{tabular}

\vspace{5mm}
\end{center}
Here the Higgs field $H$ is contained in $\Phi$ as in the $E_6$ case.
The Higgs fields $Q_L$, $\bar Q_L$, $Q_R$, and $\bar Q_R$ are
introduced only for realizing the same Kac-Moody levels of 
the three $SU(3)$ gauge groups and they do not play any other role
in the following argument.

In this model, the singlet composite operator $\bar\Phi\Phi$ plays the same
role as the FN field $\Theta$. The $D$-flatness condition for the anomalous 
$U(1)_A$ gauge symmetry is
\begin{equation}
D_A=g_A\left(\xi^2+\phi|\Phi|^2+\bar \phi|\bar \Phi|^2\right)=0,
\end{equation}
where $\xi^2$ is the parameter of the Fayet-Illiopoulos $D$-term.
Since the $D$-flatness conditions of $SU(3)_L$ and $SU(3)_R$ require
$|\VEV{\Phi}|=|\VEV{\bar \Phi}|$, the $D$-flatness condition for 
the anomalous $U(1)_A$ gauge symmetry is rewritten
\begin{equation}
D_A=g_A\left(\xi^2+(\phi+\bar\phi)|\Phi|^2\right)=0.
\end{equation}
Thus we obtain $\xi^2+(\phi+\bar \phi)|\Phi|^2=0$, namely, 
$|\VEV{\Phi}|=\left|\VEV{\bar \Phi}\right|=\xi$. 
In this case, since $\bar \Phi\Phi$ plays the same role as $\Theta$,
the unit of hierarchy becomes $\VEV{\bar \Phi\Phi}=\lambda\sim \xi^2$, which
is different from the usual case in which the FN field is just a singlet
field $\Theta$ and $\VEV{\Theta}=\xi$.
It means that even if $\xi$ has
a milder hierarchy, the unit of hierarchy becomes stronger.
Using gauge rotation and $D$-flatness condition for $SU(3)_L\times SU(3)_R$
gauge symmetry, the VEV can be taken as
\begin{equation}
\ |\VEV{\Phi}|=|\VEV{\bar \Phi }|=\left(\matrix{
0 & 0 & 0 \cr
0 & 0 & 0 \cr
0 & 0 & \lambda^{-\frac{1}{2}(\phi+\bar \phi)} \cr }\right),
\end{equation}
which breaks $SU(3)_C\times SU(3)_L\times SU(3)_R$ into
$SU(3)_C\times SU(2)_L\times SU(2)_R\times U(1)_{B-L}$.
In order to determine the VEVs of the other Higgs fields,
we examine the following superpotential
\begin{equation}
W=W_{C'}+W_{\bar C'}+W_{NV},
\end{equation}
where $W_X$ denotes the terms linear in the field $X$, which has vanishing
VEV, and $W_{NV}$ includes only the fields with non-vanishing VEVs. 
From the superpotential 
\begin{equation}
W_{NV}=\bar \Phi^3+\bar \Phi^2\bar C,
\end{equation}
$\VEV{\bar L'(\bar C)}=\VEV{\bar L^{c\prime}(\bar C)}=0$ is obtained.
The vacuum is $(\VEV{\bar S(\bar C)}=0, \VEV{\bar N^c(\bar C)}\neq 0)$ or 
$(\VEV{\bar S(\bar C)}\neq 0, \VEV{\bar N^c(\bar C)}=0)$ if 
$\VEV{\bar CC}\neq 0$. 
We are interested in the first vacuum 
$(\VEV{\bar S(\bar C)}=0, \VEV{\bar N^c(\bar C)}\neq 0)$.

The superpotential $W_{C'}$ and $W_{\bar C'}$ are given by
\begin{eqnarray}
W_{C'}&=&\lambda^{c'+\bar \phi}C'\bar \Phi
(1+\lambda^{\bar c+c}\bar CC+\lambda^{z_i+z_j}Z_iZ_j+\lambda^{z_i}Z_i)\nn \\
&&+\lambda^{c'+\bar c}C'\bar C(1+\lambda^{z_i}Z_i) \\
W_{\bar C'}&=&\lambda^{\bar c'+\phi}\bar C'\Phi
(1+\lambda^{z_i}Z_i)+\lambda^{\bar c'+c}\bar C'C(1+\lambda^{z_i}Z_i).
\end{eqnarray}
Here we neglect $(\bar \Phi\Phi)^2$ for simplicity, but the effect is critical.
After developing
the VEVs, the above interactions do not respect  $SU(3)_L\times SU(3)_R$
gauge symmetry. For example, the coefficient of $N^c(C')\bar N^c(\bar C)$
is different from that of $L(C')\bar L(\bar C)$. This is important to align
the VEVs and to give superhevay masses to these fields.
The $F$-flatness conditions 
$F_{S(C')}=F_{\bar S(\bar C')}=F_{N^c(C')}=F_{\bar N^c(\bar C')}=0$ 
determine four VEVs $\VEV{\bar CC}\sim \lambda^{-(c+\bar c)}$, 
and $\VEV{Z_i}\sim \lambda^{-z_i} (i=1,2,3)$. 
Then all the VEVs are determined by the anomalous $U(1)_A$ charges.

We now examine the mass spectrum of the Higgs sector.
The mass matrix $M_L$ for $L$ and $\bar L$ is obtained from the
interactions
\begin{equation}
\bordermatrix{
  & L^{c\prime}_\Phi & L^{c\prime}_C & \bar L'_{\bar \Phi} 
  &\bar L'_{\bar C}& L^{c\prime}_{C'} & \bar L'_{\bar C'}
  &\bar L_{\bar \Phi} & \bar L_{\bar C} &\bar L_{\bar C'} \cr
L'_\Phi& 0 & 0 & 0 & 0 & 0 & \bar C'\Phi & 0 & 0 & 
              0 \cr
L'_C & 0 & 0 & 0 & 0 & 0 & \bar C'C & 0 & 0 & 0 \cr
\bar L^{c\prime}_{\bar \Phi} & 0 & 0 & \bar \Phi^3 & \bar \Phi^2 \bar C 
              & \bar \Phi C' 
              &\bar C'\bar\Phi^2 & \bar \Phi^2\bar C & 0 & 
              \bar C'\bar C\bar \Phi \cr
\bar L^{c\prime}_{\bar C} & 0 & 0 & \bar \Phi^2\bar C & \bar\Phi\bar C^2 & 
              \bar CC' & \bar C'\bar C\bar\Phi & \bar C^2\bar\Phi & 
              \bar C^3 & \bar C'\bar C^2 \cr
L'_{C'} & 0 & 0  & C'\bar\Phi & \bar CC' & 0 &
             \bar C'C' & \bar \Phi\bar C\Phi C' & 0 &
             \bar C'\bar CC'\Phi \cr
\bar L^{c\prime}_{\bar C'} & \bar C'\Phi & \bar C'C & \bar C'\bar\Phi^2 & 
                  \bar C'\bar C\bar \Phi & \bar C'C' & \bar C'^2\bar\Phi &
                  \bar C'\bar C\bar\Phi & \bar C'\bar C^2 & \bar C'^2\bar C^2
                  \cr
L_{\Phi} & 0 & 0 & 0 & 0 & 0 & \bar C'\bar \Phi\Phi C & 0 & 0 &
               \bar C'\Phi \cr
L_{C} & 0 & 0 & 0 & 0 & 0 & \bar C'\bar\Phi C^2 & 0 & 0 & \bar C'C \cr
L_{C'} & 0 & 0 & \bar\Phi^2C'C & \bar \Phi\bar CC'C & 0 & 
            \bar C'\bar\Phi C'C & C'\bar\Phi & C'\bar C & \bar C'C' \cr
}.
\end{equation}
It is obvious that the linear combination of $\bar L'_\Phi$ and $\bar L'_C$,
and that of $L'_\Phi$ and $L'_C$ are massless and they form the doublet Higgs
fields of MSSM. 
$L_\Phi$ and $\bar L_\Phi$ are eaten by the Higgs mechanism in breaking
$SU(3)_L\times SU(3)_R$ into $SU(2)_L\times SU(2)_R\times U(1)_{B-L}$.
The mass spectrum of the remaining fields becomes 
$\lambda^{c+\bar c'}$, $\lambda^{c+\bar c'}$, $\lambda^{c+\bar c'}$,
$\lambda^{c'+\bar c}$, $\lambda^{c'+\bar c}$, $\lambda^{c'+\bar c}$, and
$\lambda^{2\bar\phi+\frac{1}{2}(\bar \phi-\phi)}$.

The mass matrix $M_E$ for $E^c$ and $\bar E^c$ is obtained from the
interactions
\begin{equation}
\bordermatrix{
  & \bar E^c_{\bar \Phi} & \bar E^c_{\bar C} & \bar E^c_{\bar C'} \cr
E^c_\Phi& 0 & 0 & \bar C'\Phi \cr
E^c_C & 0 & 0 & 0 \cr
E^c_{C'} & C'\bar\Phi & 0 & \bar C'C' \cr
}.
\end{equation}
The fields $E^c_C$ and $\bar E^c_{\bar C}$ are eaten by the Higgs mechanism
in breaking $SU(2)_R\times U(1)_{B-L}$ into $U(1)_Y$.
The mass spectrum of the remaining fields is
$\lambda^{\bar c'+\phi}$ and $\lambda^{c'+\bar \phi}$.

The mass matrix $M_{D^c}$ for the fields
$D^c, D^{c\prime}, \bar D^c, \bar D^{c\prime}$ is obtained
from the interactions
\begin{equation}
\bordermatrix{
  & \bar D^c_{\bar Q_R} & \bar D^{c\prime}_{\bar Q_R} & 
  \bar D'_{Q_L} \cr
D^c_{Q_R}         & \bar Q_RQ_R & \bar Q_R\bar\Phi Q_RC & 0 \cr
D^{c\prime}_{Q_R} & 0 & \bar Q_RQ_R & 0 \cr
\bar D'_{\bar Q_L} & \bar Q_L\bar Q_R\bar C & \bar Q_L\bar Q_R\bar\Phi
                       & \bar Q_LQ_L \cr
}.
\end{equation}
The mass spectrum becomes
$\lambda^{\bar q_r+q_r}$, $\lambda^{\bar q_r+q_r}$, and 
$\lambda^{\bar q_l+q_l}$.

The mass of the fields $Q$ and $\bar Q$
is obtained from the interaction $\bar Q_LQ_L$ as
$\lambda^{\bar q_l+q_l}$.
The mass of the fields $U^c$ and $\bar U^c$ is obtained from the interaction
$\bar Q_RQ_R$ as
$\lambda^{q_r+\bar q_r}$.

By the above argument, the mass spectrum of superheavy particles are
determined only by the anomalous $U(1)_A$ charges, 
so we can examine whether coupling unification is realized or not. 
Before going to the discussion in the next subsection, we define the reduced 
mass matrices $\bar M_I$ by getting
rid of the massless modes from the original mass matrices $M_I$.
The rank of the reduced matrices in our semi-simple model are
$\bar r_{Q}=\bar r_{U^c}=1$, $\bar r_{E^c}=2$, $\bar r_L=7$ and 
$\bar r_{D^c}=3$.
It is useful to define the effective anomalous $U(1)_A$ charges:
\begin{eqnarray}
&&x_I\equiv i+\frac{1}{2}\Delta\phi,\quad
\bar x_{\bar I}\equiv \bar i-\frac{1}{2}\Delta\phi,  
(x=l',l^{c\prime},d^{c\prime},d'), \\
&&x_I\equiv i+\Delta c-\frac{1}{2}\Delta\phi,\quad
\bar x_{\bar I}\equiv 
\bar i-\Delta c+\frac{1}{2}\Delta\phi, 
(x=l,d^c), \\
&&x_I\equiv i, \quad
\bar x_{\bar I}\equiv \bar i, 
(x=q,u^c,e^c), 
\end{eqnarray}
where $I=\Phi,C,C',Q_L,Q_R$ $(i=\phi,c,c',q_l,q_r)$, 
$\Delta\phi\equiv \frac{1}{2}(\phi-\bar\phi)$, and
$\Delta c\equiv \frac{1}{2}(c-\bar c)$.
The determinants of the reduced mass matrices are
estimated by simple sums of the effective anomalous $U(1)_A$ charges of
massive modes:
\begin{eqnarray}
\det \bar M_Q&=& \lambda^{\bar q_{\bar Q_L}+q_{Q_L}}
              =\lambda^{\bar q_l+q_l}\\
\det \bar M_{U^c}&=& \lambda^{\bar q_{\bar Q_R}+q_{Q_R}}
                  =\lambda^{\bar q_r+q_r}\\
\det \bar M_{E^c}&=&\lambda^{\bar e^c_{\bar \Phi}+e^c_{\Phi}+\bar e^c_{\bar C'}
                    +e^c_{C'}} 
                  =\lambda^{\bar\phi+\phi+\bar c'+c'}\\
\det \bar M_{D^c}&=&\lambda^{d'_{Q_L}+\bar d'_{\bar Q_L}+\bar d^c_{\bar Q_R}
                           +d^c_{Q_R}+\bar d^{c\prime}_{\bar Q_R}
                           +d^{c\prime}_{Q_R}} 
                =\lambda^{\bar q_l+q_l+2(\bar q_r+q_r)}\\
\det \bar M_L&=&\lambda^{l^{\prime}_{C}+\bar l'_{\bar \Phi}+\bar l'_{\bar C}
                 +l^{c\prime}_{C'}+\bar l'_{\bar C'}+\bar l_{\bar C}
                 +\bar l_{\bar C'}+l'_C+\bar l^{c\prime}_{\bar \Phi}
                 +\bar l^{c\prime}_{\bar C}+l'_{C'}
                 +\bar l^{c\prime}_{\bar C'}+l_C+l_{C'}} \nn \\
               &=&\lambda^{3(c+\bar c+c'+\bar c')+2\bar\phi-\Delta\phi}.
\end{eqnarray}
Then all the elements of mass matrices are estimated by simple sum of 
the effective charges 
of superheavy particles if they are not vanishing, 
and the determinants of mass matrices are
also determined by simple sum of the effective charges.
We will use this result in calculating the running gauge couplings.

\section{Coupling unification}
In this section, we apply the general discussion on the gauge coupling 
unification in Ref.\cite{maekawa3} to our scenario.
The pattern of the breaking of the gauge symmetry in our model is as follows.
At the scale $\Lambda_\Phi\sim \lambda^{-(\phi+\bar \phi)/2}$, 
$SU(3)^3$ is broken into 
$SU(3)_C\times SU(2)_L\times SU(2)_R\times U(1)_{B-L}$. 
The $SU(2)_R\times U(1)_{B-L}$ is broken into
$U(1)_Y$ at the scale $\Lambda_C\sim \lambda^{-(c+\bar c)/2}$.
We base our analysis on one loop renormalization group
equations. 
The conditions of the gauge coupling unification are given by
\begin{equation}
\alpha_3(\Lambda)=\alpha_2(\Lambda)=
\frac{5}{3}\alpha_Y(\Lambda)\equiv\alpha_1(\Lambda),
\end{equation}
where 
$\alpha_1^{-1}(\Lambda_\Phi>\mu>\Lambda_C)\equiv 
\frac{3}{5}\alpha_R^{-1}(\Lambda_\Phi>\mu>\Lambda_C)
+\frac{2}{5}\alpha_{B-L}^{-1}(\Lambda_\Phi>\mu>\Lambda_C)$,
$\alpha_1^{-1}(\mu>\Lambda_\Phi)\equiv 
\frac{4}{5}\alpha_{3R}^{-1}(\mu>\Lambda_\Phi)
+\frac{1}{5}\alpha_{3L}^{-1}(\mu>\Lambda_\Phi)$, and
$\alpha_2^{-1}(\mu>\Lambda_\Phi)\equiv 
\alpha_{3L}^{-1}(\mu>\Lambda_\Phi)
+\frac{2}{5}\alpha_{B-L}^{-1}(\mu>\Lambda_C)$.
Here $\alpha_X=\frac{g_X^2}{4\pi}$ and 
the parameters $g_X (X=3,3L,3R,2,R,B-L,Y)$ are the gauge couplings of 
$SU(3)_C$, $SU(3)_L$, $SU(3)_R$, $SU(2)_L$, $SU(2)_R$, $U(1)_{B-L}$ 
and $U(1)_Y$, respectively.

Using the fact that the three gauge couplings of the minimal SUSY standard
model (MSSM) meet at the scale $\Lambda_G\sim 2\times 10^{16}$ GeV,
the above conditions for gauge coupling unification are rewritten
\begin{eqnarray}
&&b_1\ln \left(\frac{\Lambda}{\Lambda_G}\right)
+\Sigma_I\Delta b_{1I}\ln \left(\frac{\Lambda^{\bar r_I}}{\det \bar M_I}
\right)
-\frac{12}{5}\ln \left(\frac{\Lambda}{\Lambda_C}\right)
-2\ln \left(\frac{\Lambda}{\Lambda_\Phi}\right) \label{alpha1}\\
&=&b_2\ln \left(\frac{\Lambda}{\Lambda_G}\right)
+\Sigma_I\Delta b_{2I}\ln\left(\frac{\Lambda^{\bar r_I}}{\det \bar M_I}
\right) -2\ln \left(\frac{\Lambda}{\Lambda_\Phi}\right) \label{alpha2}\\
&=&b_3\ln \left(\frac{\Lambda_A}{\Lambda_G}\right)
+\Sigma_I\Delta b_{3I}\ln\left(\frac{\Lambda_A^{\bar r_I}}{\det \bar M_I}
\right),
\end{eqnarray}
where $(b_1,b_2,b_3)=(33/5,1,-3)$ are the 
renormalization group coefficients
for MSSM
and $\Delta b_{aI}(a=1,2,3)$ are the corrections to the coefficients 
from the massive fields $I=Q+\bar Q,U^c+\bar U^c, E^c+\bar E^c, D^c+\bar D^c$,
and $L+\bar L$.
The second last term in Eq. (\ref{alpha1}) is from the breaking 
$SU(2)_R\times U(1)_{B-L}\rightarrow U(1)_Y$ by the VEV $\VEV{C}$,
and the last terms in Eqs. (\ref{alpha1}), (\ref{alpha2}) are from the
breaking 
$SU(3)_L\times SU(3)_R\rightarrow SU(2)_L\times SU(2)_R\times U(1)_{B-L}$
by the VEV $\VEV{\Phi}$. 
Since all the mass matrices and the symmetry breaking scales appearing 
in the above conditions are determined by the anomalous $U(1)_A$ charges,
these conditions can be translated to the constraint on the effective
charge and the cutoff scale,
\begin{eqnarray}
\alpha_1(\Lambda)=\alpha_2(\Lambda)&\rightarrow &
\Lambda\sim \Lambda_G\lambda^{-\frac{1}{28}(5\bar\phi-\phi+6(c+\bar c))},\\
\alpha_2(\Lambda)=\alpha_3(\Lambda)&\rightarrow &
\Lambda\sim \Lambda_G\lambda^{\frac{1}{8}
(7\bar\phi+\phi+6(c+\bar c+c'+\bar c'))},\\
\alpha_1(\Lambda)=\alpha_3(\Lambda)&\rightarrow &
\Lambda\sim \Lambda_G\lambda^{\frac{1}{96}
(25\bar\phi+19\phi+30(c+\bar c)+18(c'+\bar c')}.
\end{eqnarray}
A naive calculation leads to the relation between the charges,
\begin{equation}
59\bar\phi+41\phi+42(c+\bar c)+54(c'+\bar c')\sim 0,
\end{equation}
which is difficult to satisfy in our scenario.
However, careful calculation shows that gauge coupling unification
is possible, though somewhat larger ambiguities of order one coefficients are
required than in a simple group unification.
Actually with the typical charge assignment in Table I, the coupling
unification is realized as in Fig. 1, using the ambiguities of
order one coefficients $\lambda\leq y\leq\lambda^{-1}$. 
\begin{figure}[htb]
\begin{center}
\leavevmode
\epsfxsize=110mm
\put(300,50){{\large $\bf{\log \mu (GeV)}$}}
\put(0,260){{\Large $\bf{\alpha^{-1}}$}}
\put(29,240){$\alpha_1^{-1}$}
\put(31,150){$\alpha_2^{-1}$}
\put(31,90){$\alpha_3^{-1}$}
\epsfbox{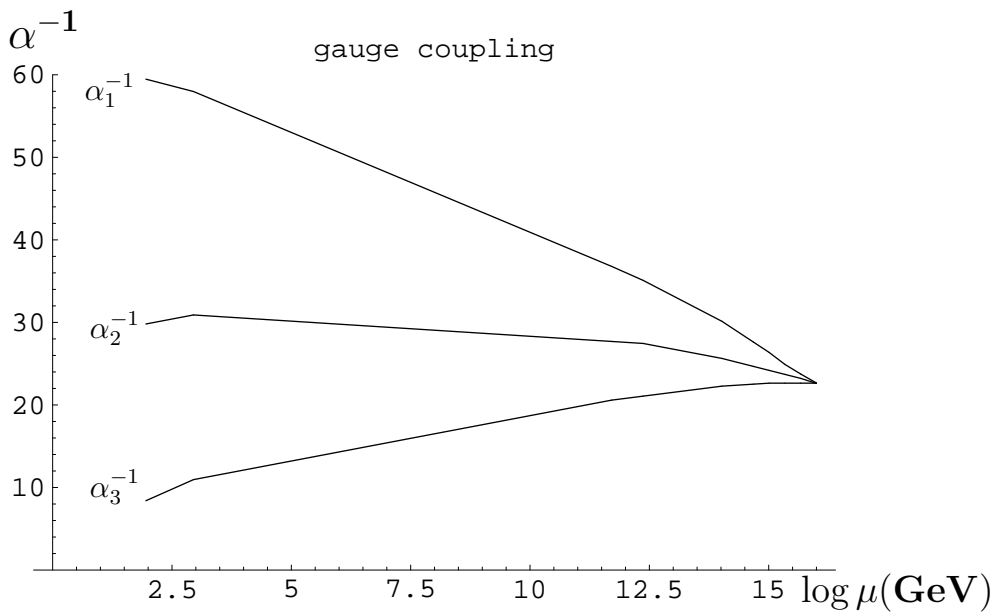}
\vspace{-2cm}
\caption{
Here we adopt $\lambda=0.22$, 
$\alpha_1^{-1}(M_Z)=59.47$, $\alpha_2^{-1}(M_Z)=29.81$,
$\alpha_3^{-1}(M_Z)=8.40$, the SUSY breaking scale $m_{SB}\sim 1$ TeV and
the anomalous $U(1)_A$ charges: 
$(\psi_1,\psi_2,\psi_3)=(9/2,7/2,3/2)$, $\phi=-3$, $\bar \phi=2$, $c=-3$,
$\bar c=1$, $z=-1$, $c'=0$ and $\bar c'=4$.
Using the ambiguities of coefficients $\lambda\leq y\leq \lambda^{-1}$,
three gauge couplings meet at around 
$10^{16}$ GeV. 
}
\label{fig_1}
\end{center}
\end{figure}

Note that the unified gauge coupling can remain in the perturbative region
in $SU(3)^3$ model, which is different from the $E_6$ model.
This is because the Higgs sector in $SU(3)^3$ unification is much simper
than that in $E_6$ unification, since we do not have to introduce
adjoint Higgses to realize the doublet-triplet splitting.

The cutoff scale tends to be lower than the Planck scale. 
Indeed, the cutoff is taken as $10^{16}$ GeV in Fig. 1. 
Since the cutoff scale is so low, we have to take care of
proton decay via dimension five operators
\cite{SY}, which
are obtained from 
\begin{equation}
\lambda^{\psi_i+\psi_j+\psi_k+\psi_l+\bar \phi}
\Psi_i\Psi_j\Psi_k\Psi_l\bar \Phi
\end{equation}
by developing the VEV 
$\VEV{\bar \Phi}\sim \lambda^{-\frac{1}{2}(\phi+\bar \phi)}$.
The coefficients are suppressed not only by the usual small Yukawa factor 
but also by the suppression factor 
$\lambda^{4n+\frac{1}{2}(\bar \phi-\phi}=\lambda^{8.5}$.
Even if we take the cutoff $\Lambda\sim 10^{16}$ GeV, the `effective'
colored Higgs mass is around $\lambda^{-8.5}\Lambda\sim 10^{22}$
GeV, which is much larger than the experimental bound of $10^{18}$ GeV.
Thus, proton decay via dimension five operator is adequately suppressed. 

On the other hand, proton decay $p\rightarrow e^+\pi^0$
via dimension six operators from the K\"ahler potential
\begin{equation}
K=\frac{1}{\Lambda^2}\Psi_1^\dagger \Psi_1\Psi_1^\dagger\Psi_1,
\end{equation}
which are allowed by the symmetry in our
scenario by taking the unification scale $\Lambda_U$ as the cutoff $\Lambda$,
may be seen in future experiments. If we roughly estimate the lifetime 
of proton
using the formula in Ref.~\cite{hisano} and the recent result of 
the lattice calculation for the hadron matrix element parameter
$\alpha$\cite{lattice}, we find
\begin{equation}
\tau_p(p\rightarrow e^+\pi^0)\sim 4.5\times 10^{34}\left(\frac{\Lambda}
{10^{16}{\rm GeV}}\right)^4
\left(\frac{0.015({\rm GeV})^3}{\alpha}\right)^2  {\rm years}.
\end{equation}
This estimate, albeit a rough one, provides a strong motivation for
continuing the proton decay search.

\section{Discussion and Summary}
Besides $SU(3)^3$, $E_6$ has the other maximal semi-simple subgroups
$SU(6)\times SU(2)_L$ and $SU(6)\times SU(2)_R$\cite{Matsuoka}.
The matter sector can be applied to these subgroups in a 
straightforward way. However, in the Higgs sector, it is difficult 
to realize
the situation in which only one pair of doublet Higgs is massless.
It is difficult to make the partner of the doublet Higgs
massive, while keeping the latter massless. On the contrary,
in $SU(3)^3$ gauge symmetry, since the partners of the doublet Higgs $L$,
$E^c$ and $N^c$ are absorbed by the Higgs mechanism, one pair of doublet
Higgs can be massless. 

In the typical charge assignment, the charges of the matter sector
respect the $E_6$ symmetry, while those of the Higgs sector do not.
It is difficult to respect $E_6$ symmetry in the Higgs sector without
additional massless fields other than the fields in the MSSM. 

By introducing singlet fields, we can build models with integer 
Kac-Moody level. For example, in addition to the fields in Table I,
we introduce a singlet field with charge 10, one with charge -8, 
43 singlet fields with charge 3/2, and 62 singlet fields with charge 1/2.
Then using the relation
\footnote
{ 
$C_a\equiv\Tr_{G_a} T(R)\,Q_A$.  Here
$T(R)$ is the Dynkin index of the representation $R$, and we use the
convention in which $T(\mbox{fundamental rep.}) = 1/2$.
}
\begin{equation}
\frac{C_a}{k_a}
\,= \,\frac{1}{3k_A}\tr {Q_A}^3 
\,= \,\frac{1}{24}\tr Q_A,
\end{equation}
where $k_A$ and $k_a$ are Kac-Moody levels of $U(1)_A$ and $SU(3)_a$
($a=C,L,R$), 
these Kac-Moody levels can be calculated as
\begin{equation}
k_A=4,\ k_C=k_L=k_R=2.
\end{equation}
Note that introducing the singlets with charge 10 and -8, the $\mu$ problem
is solved by the mechanism proposed in Ref. \cite{maekawa2}. 

In our model the difference between the mass matrices of down-type quark
and the charged leptons are realized because the matrices are
from different Yukawa interactions. However, if this model
is regarded as the low energy theory of $E_8\times E_8$ heterotic 
superstring theory, we have to break the gauge symmetry $E_6$
into $SU(3)^3$. Since the matter sector respects $E_6$ symmetry,
it is natural to expect that the Yukawa interactions also respect it.
In order to realize the different Yukawa interactions,
we have to implement the breaking. In the brane world scenario, there is an
interesting mechanism to break the gauge symmetry\cite{extra}.
However, it seems difficult to realize the breaking in the Yukawa 
coupling of matter which resides on the brane. To enforce the $E_6$ breaking
some of the matter must be in the bulk, where the $E_6$ gauge symmetry
is not respected.

In this paper, we have proposed a realistic semi-simple unified theory
with $SU(3)^3$ gauge group. Since generic interactions have been introduced,
we can define the model by the anomalous $U(1)_A$ charges.
Large neutrino mixing angles can be realized in the model.
Moreover, the FCNC process is automatically suppressed. 
The half integer charges of the matter sector automatically play the 
role of $R$-parity. The model has the same matter structure as in
$E_6$ model\cite{BM}, but different and simpler Higgs structure. 
Actually we do not need the adjoint Higgs fields to realize doublet-triplet
splitting. And the gauge couplings at the cutoff scale can 
be in the perturbative region.
Note that in the $SU(3)^3$ model, 
in contrast to $SU(5)$, $SO(10)$ or $E_6$, the lightest magnetic 
monopole carries three (instead of one) quanta
of Dirac magnetic charge\cite{shafi4}. This is readily seen by noting 
that one is allowed, in principle, to include non- bifundamental 
vectorlike representations such as $(\bf{1,3,1})$ + $({\bf 1,\bar 3,1})$
that, despite their color singlet nature, carry
fractional(e/3) electric charge. The Dirac quantization then requires that
the corresponding magnetic charge has three units. The number density of
primordial $SU(3)^3$ monopoles depends, of course, 
on the underlying cosmological
scenario, and should not exceed the nominal Parker bound of about 
$10^{-16} {\rm cm}^{-2} {\rm s}^{-1} {\rm sr}^{-1}$. 
The discovery of magnetic monopoles would 
be a truly remarkable event, and measurement of their magnetic charge would 
allow us to distinguish
between a variety of unified gauge theories.

It would be interesting to extend the approach presented here
to other semi-simple unification schemes. For instance, the gauge symmetry
$SU(3)^3$ with three {\bf 27}'s of $E_6$ can be embedded, in principle,  in
$SU(4)\times SU(3)\times SU(3)$\cite{shafi5} which could be worth pursuing. 

\section{Acknowledgement}
We would like to thank the organizers of the workshops NOON2001 at
Kashiwa and post NOON at Kyoto, where this collaboration began.
Q.S. is supported in part by the DOE under Grant No. DE-FG02-91ER40626.

\end{document}